\documentclass[review]{elsarticle}

\usepackage{amssymb}

\usepackage{hyperref}

\journal{Nuclear Instruments and Methods in Physics Research A~}

\begin{document}

\begin{frontmatter}

\title{UCGretina \textsc{geant4} Simulation of the GRETINA Gamma-Ray
  Energy Tracking Array}

\author[ursinusaddress]{L. A. Riley\corref{correspondingauthor}}
\cortext[correspondingauthor]{Corresponding author}
\ead{lriley@ursinus.edu}

\author[nscladdress]{D. Weisshaar}

\author[lbnladdress]{H. L. Crawford}

\author[ursinusaddress]{M. L. Agiorgousis}

\author[lbnladdress]{C. M. Campbell}

\author[lbnladdress]{M. Cromaz}

\author[lbnladdress]{P. Fallon}

\author[nscladdress,msuaddress]{A. Gade}

\author[ursinusaddress]{S. D. Gregory}

\author[ursinusaddress]{E. B. Haldeman}

\author[ursinusaddress]{L. R. Jarvis}

\author[ursinusaddress]{E. D. Lawson-John}

\author[ursinusaddress]{B. Roberts}

\author[ursinusaddress]{B. V. Sadler}

\author[ursinusaddress]{C. G. Stine}

\address[ursinusaddress]{Ursinus College, Collegeville, PA, USA}

\address[nscladdress]{National Superconducting Cyclotron Laboratory,
  Michigan State University, East Lansing, MI 48824, USA}

\address[msuaddress]{Department of Physics and Astronomy, Michigan
  State University, East Lansing, Michigan 48824, USA}

\address[lbnladdress]{Lawrence Berkeley National Laboratory, Berkeley,
  CA 94720, USA}

\begin{abstract}
UCGretina, a \textsc{geant4} simulation of the GRETINA gamma-ray
tracking array of highly-segmented high-purity germanium detectors is
described. We have developed a model of the array, in particular of
the Quad Module and the capsules, that gives good 
agreement between simulated and measured photopeak efficiencies over a
broad range of gamma-ray energies and reproduces the shape of the
measured Compton continuum. Both of these features are needed in order
to accurately extract gamma-ray yields from spectra collected 
in in-beam gamma-ray spectroscopy measurements with beams
traveling at $v/c \gtrsim 0.3$ at the National Superconducting
Cyclotron Laboratory and the Facility for Rare Isotope Beams. In the
process of developing the model, we determined that millimeter-scale
layers of passive germanium surrounding the active volumes of the
simulated crystals must be included in order to reproduce measured
photopeak efficiencies. We adopted a simple model of effective passive
layers and developed heuristic methods of determining passive-layer  
thicknesses by comparison of simulations and measurements for a single
crystal and for the full array. Prospects for future development of
the model are discussed.
\end{abstract}


\end{frontmatter}

%\linenumbers

\section{Introduction}

The Ursinus College Gretina simulation code (UCGretina)
is a \textsc{geant4}~\cite{Geant4, Geant4b} simulation code used in
the planning and analysis of measurements (see for example
Refs.~\cite{Ril14, Lau17, Gad19a, Gad19b})
%~\cite{Ril14, Str14, Ril16, Lun16, Lau17, Ril17, Gad19a, Gad19b,
% Elm19})
made with the
\textbf{G}amma-\textbf{R}ay \textbf{E}nergy \textbf{T}racking
\textbf{I}n-beam \textbf{N}uclear \textbf{A}rray
(GRETINA)~\cite{GRETINA, GRETINA2}, the initial 
stage of the \textbf{G}amma-\textbf{R}ay \textbf{E}nergy
\textbf{T}racking \textbf{A}rray (GRETA)~\cite{Bea03} now under 
construction. A primary application of UCGretina is in fitting the
simulated response of GRETINA to measured Doppler-reconstructed
gamma-ray spectra collected in measurements of rare isotope beams
traveling at $v/c \gtrsim 0.3$ in order to determine gamma-ray
yields. Simulations are of particular importance in measurements
involving significant scattering and absorption of gamma rays by
experimental apparatus surrounding the target material -- a gas or
liquid target cell, or a compact charged-particle array, for
example. Reproducing the measured response of the array requires an
accurate model of all components of GRETINA, including not only the
active HPGe detectors and well-understood passive materials, but also
the ill-defined effective passive germanium layers associated with the
HPGe detector contacts and surfaces.

In the present work, we begin, in Section~\ref{sec:code}, by
describing the simulation models of GRETINA and the GRETINA scanning
table implemented in the UCGretina code with an emphasis on passive
material included in the models. In Sections~\ref{sec:pencil} and
\ref{sec:sources}, we describe pencil-beam source and flood-source
measurements used to constrain and validate the model of the array
used in the code.

In Sections~\ref{sec:passive_single} and~\ref{sec:full}, we describe
heuristic methods for determining optimal average effective
thicknesses for the passive germanium layers of both a single crystal
and for the full array. Measured photopeak efficiencies and the
Compton continuum are compared to the simulations to assess how well
these effective passive-layers reproduce the detector response.  We
characterize these methods as \textit{heuristic} and the passive
germanium layers in the model as \textit{effective}, due to the simple
passive-layer geometry assumed and due to the fact that no attempt is
made here to model processes such as partial charge collection along
passive-layer boundaries.  Passive layers in high-purity germanium
detectors, in which charge collection is inhibited, are thought to
arise due to weaker electric fields near the crystal surface caused by
the presence of surface charges attributed to the chemical passivation
treatment of the outer surfaces of the crystal~\cite{Ebe08}.
Non-uniformities in the field are more pronounced in single-ended
coaxial crystals like those used in GRETINA. Passive-layer geometries
have been shown to be dependent on the energy of the incident gamma
ray, and the underlying physics is not well understood (see
Ref.~\cite{Mai16} and works cited therein). Attempting to explore the
detector physics of passive layers at HPGe surfaces is beyond the
scope of the present work, and we focus instead on constraining
simpler effective passive layers to reproduce the overall array
response.

We find that by varying the various effective passive-layer
thicknesses, we cannot find a parameter set which reproduces all
measured aspects of GRETINA in a single simulation model. Problematic
is the low-energy range up to about 250~keV, which is mostly affected
by the choice of the thickness of the outer passive layer at the front
surface of the crystals. In addition, we find that substantial passive
material located behind and adjacent to the active detector volumes,
such as the aluminum mounting shell, is required to obtain good
agreement with the measured Compton continuum. 

\section{The UCGretina Code}
\label{sec:code}

\begin{figure}
  \begin{center}
    \begin{tabular}{cc}
      \scalebox{0.18}{
        \includegraphics{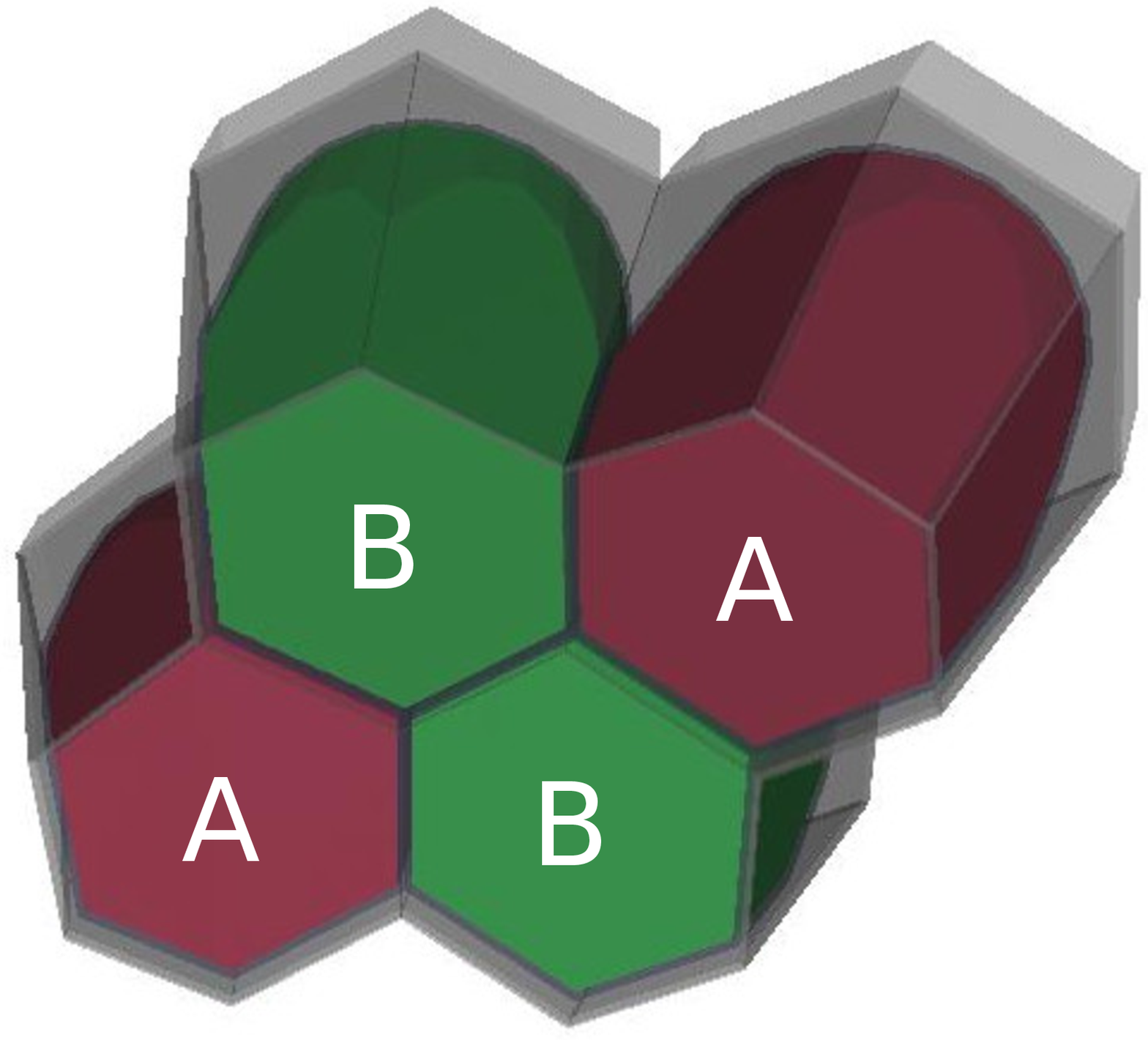}
      }
      &
      \scalebox{0.18}{
        \includegraphics{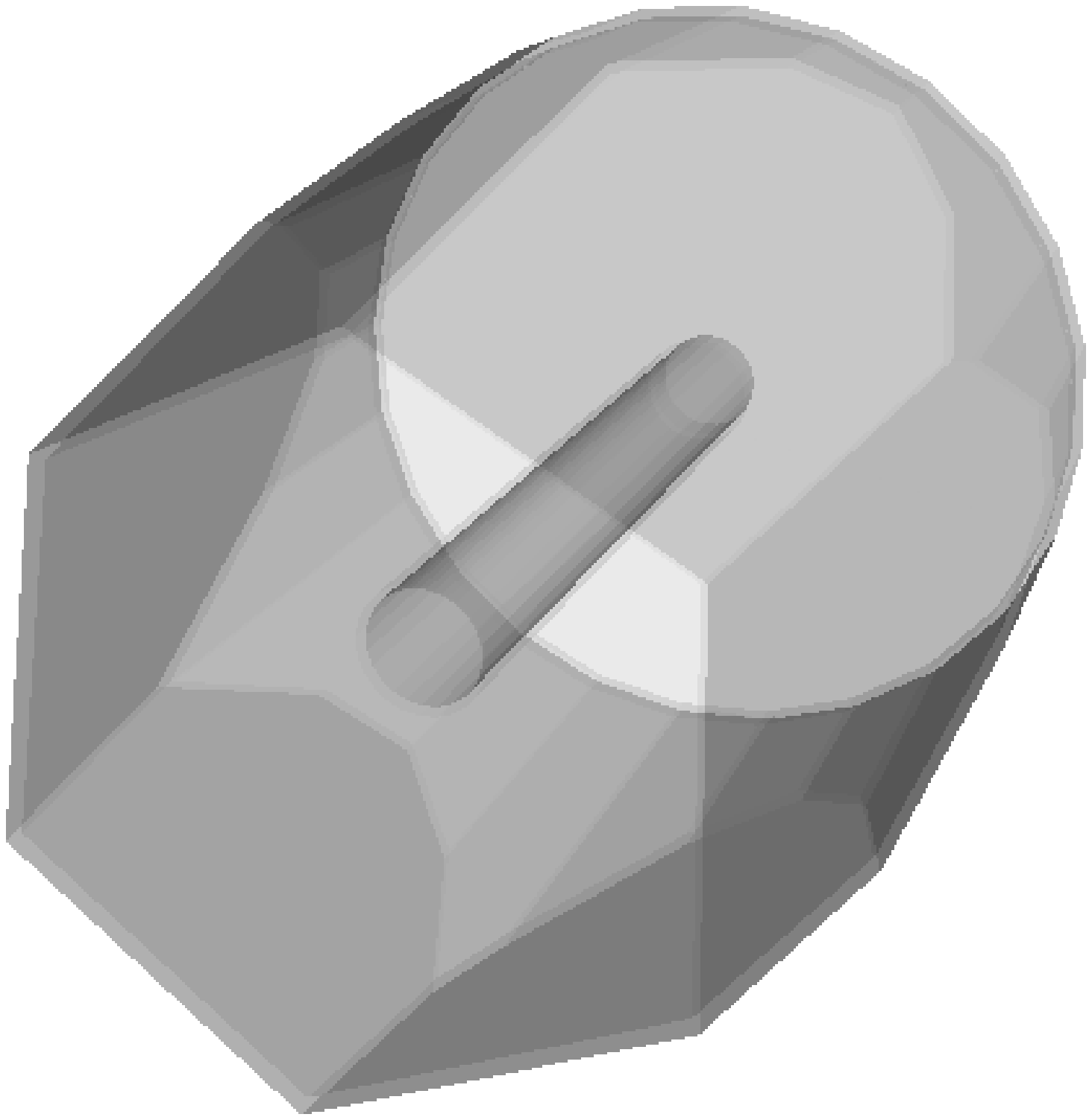}
      }\\
      \textbf{(a)} & \textbf{(b)}
    \end{tabular}
    \caption{\label{fig:ucg_quad} Graphical renderings of a single
      GRETINA quad showing (a) the crystals, labeled by type,
      with aluminum capsules and endcaps and (b) the coaxial,
      back, and outer passive layers surrounding the active detector
      volume of a Type a crystal.}
  \end{center}
\end{figure}

The UCGretina application is built using version 4.10.05 of the
\textsc{geant4} toolkit~\cite{Geant4,Geant4b}. The
\texttt{G4EmStandardPhysics\_option4} physics 
list is used, which includes the collection of models best suited to
low-energy physics applications. In simulations of in-beam
measurements, beam nuclei are tracked through the target and undergo gamma
decay in flight. The \textsc{geant4} toolkit manages
discrete gamma emission by excited nuclei with the nuclear
de-excitation module. By default, level scheme data are
drawn from the \textsc{geant4} \texttt{PhotonEvaporation} data file
based on the Evaluated Nuclear Structure Data File
(ENSDF)~\cite{ENSDF}. UCGretina also supports user-specified
level-schemes. Beta decay, electron capture, and alpha decay can be
simulated with the radioactive decay module, using
decay data from the \textsc{geant4} \texttt{RadioactiveDecay} data
file based on ENSDF.

In UCGretina, models of the germanium crystals are built and arranged,
along with aluminum capsules surrounding each crystal and aluminum
endcaps housing each cluster of four crystals, using a 
version of the detector geometry specification of the simulation
code~\cite{Far04} developed for the Advanced GAmma Tracking Array
(AGATA)~\cite{Akk12} modified to include an outer Ge passive layer. A
graphical rendering of the model of a single cluster of four crystals
-- called a ``quad'' -- is shown in panel (a) of
Figure~\ref{fig:ucg_quad}. The overall shape of each crystal is 
defined as the logical union of a cylinder with a six-sided convex
polyhedron. There are two slightly different crystal shapes, types A
and B, labeled in Figure~\ref{fig:ucg_quad}(a). Each crystal also 
contains a central core contact --- a cylindrical, coaxial void
that extends from the back surface to 15~mm from the front face. The
original model included passive Ge layers at the back of the
crystal and surrounding the coaxial central contact. We have added an
outer passive Ge layer to the model and show in
Sections~\ref{sec:passive_single} and~\ref{sec:full} that it
substantially improves the agreement of simulations with measured
photopeak efficiencies. The passive layers of a Type A crystal are
illustrated in Figure~\ref{fig:ucg_quad}(b).

\begin{figure}
  \begin{center}
    \begin{tabular}{c}
      \scalebox{0.31}{
        \includegraphics{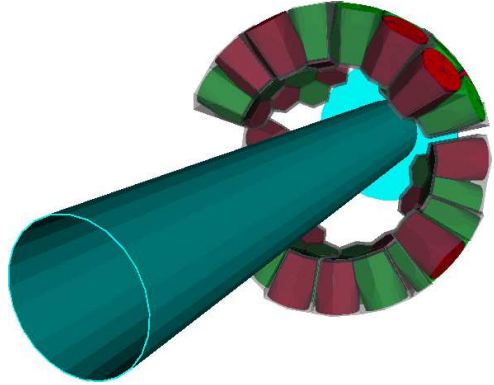}
      }\\
      \textbf{(a)}\\
      \\
      \scalebox{0.35}{
        \includegraphics{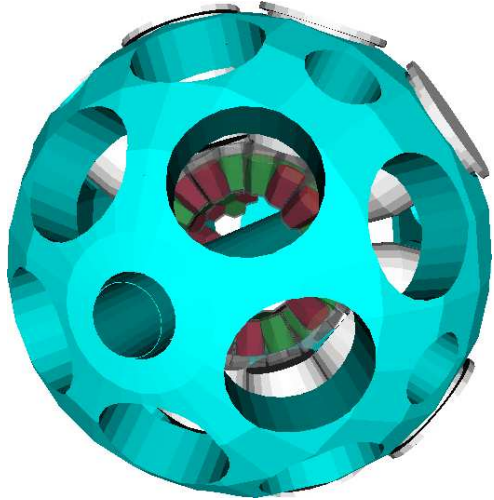}
      }\\
      \textbf{(b)}
    \end{tabular}
    \caption{\label{fig:ucg_full} Graphical renderings of the full
      GRETINA array (a) including the crystals, capsules, end caps,
      and beam pipe and (b) including additional passive material
      that can scatter particles into the active detector
      volumes.} 
  \end{center}
\end{figure}

Graphical renderings of the main 12-quad configuration of GRETINA
used in the third GRETINA campaign in 2019-2020 at the National
Superconducting Cyclotron Laboratory (NSCL) appear in
Figure~\ref{fig:ucg_full}. Panel (a) shows a minimal model of the
array needed to reproduce measured 
photopeak efficiencies, which includes all components of the apparatus
that fall at least partially between the target (source) and the
active detector volumes. The target is located at the center of
GRETINA and is obscured by the beam pipe in Figure~\ref{fig:ucg_full}.
The model in panel (b) 
includes additional passive material that can scatter gamma rays into
the active detector volumes, including simple models of the
detector cryostat and liquid nitrogen dewars behind the crystals, the
GRETINA mounting shell, and the gate valve and the housing of the
quadrupole magnet at the entrance of the S800 Magnetic
Spectrograph~\cite{S800}, not visible in
Figure~\ref{fig:ucg_full}(b). Also included, but not shown, is an
aluminum sphere of inner radius 0.95~m surrounding the entire array, which
accounts for scattering from objects outside of the mounting
shell. For reference, the outer radius of the mounding shell is
0.64~m. We demonstrate in Section~\ref{sec:Compton} that this
additional passive material is needed in order to reproduce the
measured Compton continuum. 

\begin{figure}
  \begin{center}
    \scalebox{0.4}{
      \includegraphics{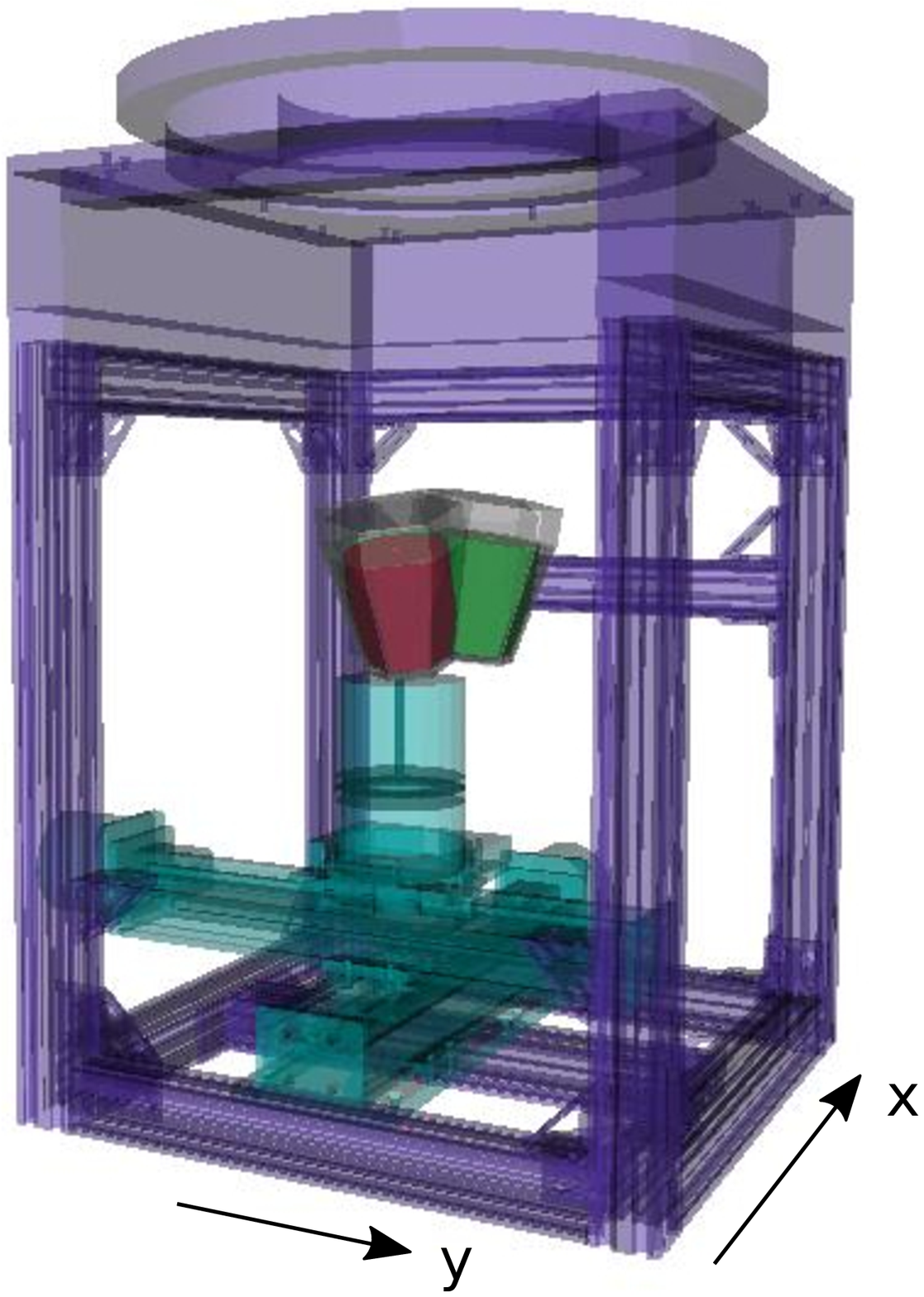}
    }
    \caption{\label{fig:scan} Graphical rendering of the simulation
      of a quad in the GRETINA scanning table.} 
  \end{center}
\end{figure}

Measurements made with a collimated (so-called ``pencil beam'') source
using the GRETINA scanning table at the Lawrence Berkeley National
Laboratory (LBNL) are presented in Section~\ref{sec:pencil} and
compared with simulations in Section~\ref{sec:coaxial_single}.  A
model of the scanning table has been implemented in UCGretina, using
the CADMesh package~\cite{CADMesh} to import CAD designs into
\textsc{geant4}. A rendering of the model is shown in
Figure~\ref{fig:scan}. It includes more passive material than is
strictly needed for the simulations of the pencil-beam photopeak
efficiencies presented in the present work, which are only affected by
material directly between the source and the active volumes
of the crystals. 

\section{Pencil-Beam Source Measurements}
\label{sec:pencil}

Pencil-beam source measurements of a type-A GRETINA crystal in a quad
detector module were made at LBNL using the GRETINA scanning table. A 
1~mCi $^{137}$Cs source housed in a Hevimet (W 90\%, Ni 6\%, Cu 4\%)
collimator with a 1~mm 
diameter opening and an 87~mm long bore was used to produce a beam
with angular opening of 11~mrad. The quad was mounted in the scanning
table with its axis oriented vertically, facing downward as
illustrated in Figure~\ref{fig:scan}. The collimator was mounted
on two perpendicular linear stages, positioned independently by ball
screws driven by stepper motors.  Four scans of measurements with
points spaced along the y-axis were made, separated by 5~mm along the
x-axis of the scanning table, in a horizontal plane perpendicular to
the quad axis. A fifth scan was made in the region of the central
contact, at a fixed y-position with points spaced along the x-axis. 
The granularity of the data points within each scan was 2~mm near the
central contact and 4 mm for the remainder of the points. Data was
collected at 20 points per scan for a total of 100 measurement points.
These measurements were used in the optimization of the coaxial
passive-layer thickness of a single crystal presented in
Section~\ref{sec:coaxial_single}.

\section{Flood-Source Measurements}
\label{sec:sources}

The GRETINA array was installed in its main 12-quad configuration at
the NSCL with four quads mounted at 58$^\circ$ with respect to the
beam axis and the remaining eight quads at 90$^\circ$ as shown in the
rendering of Figure~\ref{fig:ucg_full}.  Sources were placed at the
center of the array, mounted on a G10 fiberglass laminate ring held in
place by a target cradle in the 6 inch diameter aluminum beam pipe.
Absolute photopeak efficiency measurements were made with $^{22}$Na,
$^{57}$Co, $^{133}$Ba, $^{152}$Eu, and $^{241}$Am sources with known
activities. The measured and simulated Compton continuums in this
spectrum are compared in Section~\ref{sec:Compton}. Relative
efficiency measurements of 
$^{56}$Co and $^{226}$Ra were scaled to fit measured absolute
efficiencies below 1836~keV in order to extend the measured photopeak
efficiencies to 3548~keV. Low-energy thresholds for all crystals were
set below 50~keV. The methods used are described in greater detail in
Ref.~\cite{GRETINA2}. The full set of photopeak efficiency measurements
covers energies from 53~keV to 3548~keV. These photopeak efficiency
measurements were used in the optimization of the back and outer
passive-layer thicknesses used in the UCGretina model described in
Sections~\ref{sec:passive_single} and~\ref{sec:passive_full}.
Additionally, absolute efficiencies were measured using coincidence
measurements of $^{60}$Co and $^{88}$Y sources with a LaBr$_3$:Ce
scintillation detector centered in the beam line just upstream of the
source. A virtually background-free spectrum of the response of
GRETINA to the 1172~keV gamma ray in $^{60}$Ni in coincidence with a
software gate on the 1332~keV gamma ray in the LaBr$_3$:Ce detector
was also collected. This spectrum is critical to examining the
complete response of the array, with comparison of the simulated and
measured Compton continuums in Section~\ref{sec:Compton}.

\section{Optimal Passive Layers: Single Crystal}
\label{sec:passive_single}

Here, we describe a heuristic method for determining optimal effective
passive-layer thicknesses in the UCGretina crystal model for the
single crystal with which the pencil-beam scans
described in Section~\ref{sec:pencil} were made. 
We first determine a tentative coaxial passive-layer thickness using 
pencil-beam scans as described in
Section~\ref{sec:coaxial_single}.
Then, the corresponding back and outer passive-layer thicknesses are
determined as described in Section~\ref{sec:back_outer_single} using
the photopeak efficiency measurements described in
Section~\ref{sec:sources}. The back passive-layer thickness is
constrained using the fraction of photopeak events involving at least
one gamma-ray interaction point in the back slice of the crystal, and
the outer passive-layer thicknesses is determined using the 
photopeak efficiencies of the full crystal. Finally, these 
steps are repeated iteratively until the process converges on an
optimal set of effective passive-layer thicknesses.

In the comparisons of simulations with measurements in the present
work, we determine best-fit parameter values using a weighted
least-squares method, minimizing the Neymann $\chi^2$, which for
statistically independent observations is given by~\cite{MINUIT}
\begin{equation}
  \chi^2 = \sum_i \frac{(N_i - N^\mathrm{sim}_i)^2}{N_i}
\end{equation}
where the $N_i$ are measured and the $N^\mathrm{sim}_i$ are simulated
photopeak counts. 

\subsection{Coaxial Passive Layer}
\label{sec:coaxial_single}

\begin{figure}
  \begin{center}
    \scalebox{0.80}{
      \includegraphics{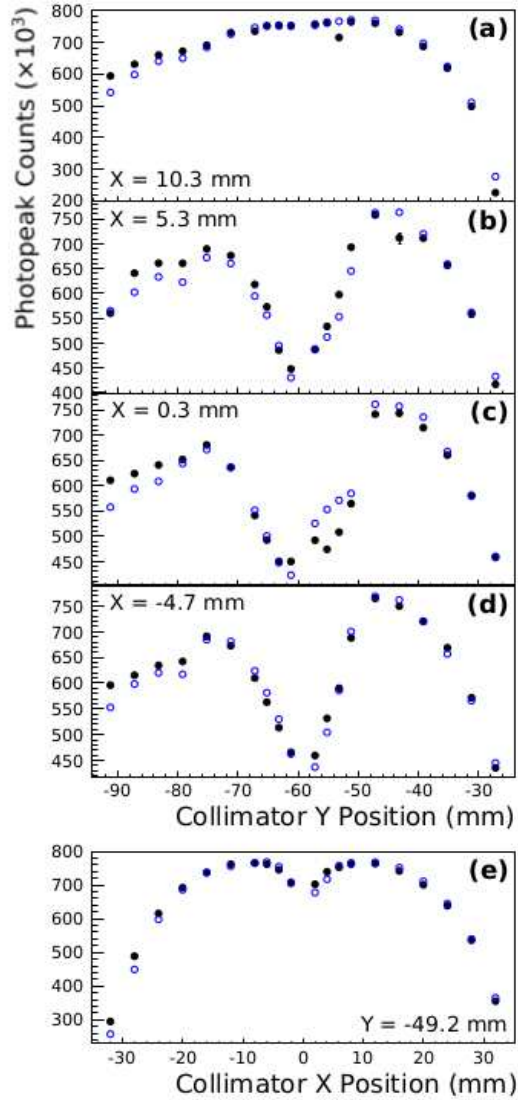}
    }
    \caption{\label{fig:pencil} Measured (full circles) and simulated
      (open circles) photopeak counts from pencil-beam scans of
      a type-A crystal in a GRETINA quad. Four scans were made along
      the scanning-table y axis oriented toward the center of the quad
      (panels a-d) and one pencil beam scan along the scanning-table x
      axis, across the central contact of the crystal.
      The optimal 2.09~mm coaxial, 2.8~mm back, and 0.40~mm
      outer effective passive-layer thicknesses were used to produce
      the simulated points.}  
  \end{center}
\end{figure}

Simulated photopeak yields from pencil-beam scans crossing the region
of the central contact are highly sensitive to the thickness of the
coaxial passive layer surrounding the central contact assumed in the
crystal model. Plots of measured and simulated photopeak yields
vs. collimator position for the five pencil-beam scans of a type-A
crystal in a GRETINA quad described in Section~\ref{sec:pencil} are
shown in Figure~\ref{fig:pencil}.  The response of the detector to
each measurement in the pencil beam scans was simulated assuming
values of the coaxial passive-layer thickness from 1~mm to 4~mm in
0.5~mm steps. For each passive-layer thickness, the simulations were
scaled to the measured photopeak yields in the full set of five
pencil-beam scans.

Horizontal offsets of the simulated collimator position and the
rotation of the quad about its axis were varied to optimize the
overall fit. This process yielded a 0.9 mm offset along the scanning
table Y axis and a 1$^\circ$ rotation of the quad relative to the
nominal simulation model. Once the offset and rotation were fixed, the
only remaining free parameter in the fit was the scaling of the
simulated yields. A common scaling was applied to all five scans in 
the fitting process. The corresponding simulated photopeak yields
appear as open circles in Figure~\ref{fig:pencil}. The final best-fit
effective coaxial passive-layer thickness was found, via $\chi^2$
minimization, to be 2.091~mm with a 95\% confidence interval of
0.012~mm. This is a purely statistical uncertainty within the very
limited parameter space of the model. For example, it does not account
for variations of actual crystal dimensions within the tolerances
reported in manufacturer drawings, and it does not reflect the
precision with which passive-layer thicknesses can be determined.

\subsection{Back and Outer Passive Layers}
\label{sec:back_outer_single}

\begin{figure}
  \begin{center}
    \scalebox{0.5}{
      \includegraphics{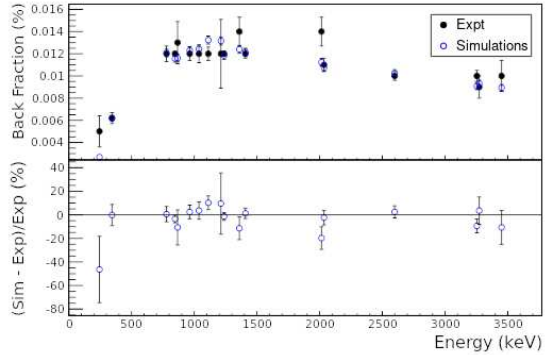}
    }
    \caption{\label{fig:Q4C4_back} (Top) Measured (full circles) and
      simulated (open circles) fraction of photopeak events in a
      type-A crystal in a GRETINA quad with at least one interaction
      point in the back slice of the crystal using the optimal
      effective passive-layer thicknesses in the simulations. (Bottom)
      Relative discrepancies between simulations and measurements.}
  \end{center}
\end{figure}

The back and outer passive-layer thicknesses are constrained using the
photopeak efficiency measurements described in
Section~\ref{sec:sources}. 
The fraction of photopeak events in which at
least one gamma-ray interaction point registered within one of the six
segments at the back of the crystal are sensitive to the thickness of
the back passive layer assumed in the crystal model. This 
``back fraction'' measured with a type-A crystal in a GRETINA quad is
compared with simulations at 17 energies in the range 
344~keV to 3548~keV in Figure~\ref{fig:Q4C4_back}. Energies below
344~keV were excluded due to the low probability of the lower-energy
gamma rays reaching the back slice of the crystals.
Simulations assuming back passive-layer thicknesses between 1.5~mm and
4~mm with step size 0.5 mm were made. 
A best-fit effective back passive-layer thickness of 2.8~mm with a
95\% confidence interval of 0.4~mm was found via $\chi^2$
minimization. This is a purely statistical 
uncertainty within the limited parameter space of the model and does
not reflect the precision with which passive-layer thicknesses can be
determined. Back fractions simulated
with the resulting optimal effective back passive-layer thickness are
shown as open circles in the upper panel of
Figure~\ref{fig:Q4C4_back}. 

\begin{figure}
  \begin{center}
    \scalebox{0.5}{
      \includegraphics{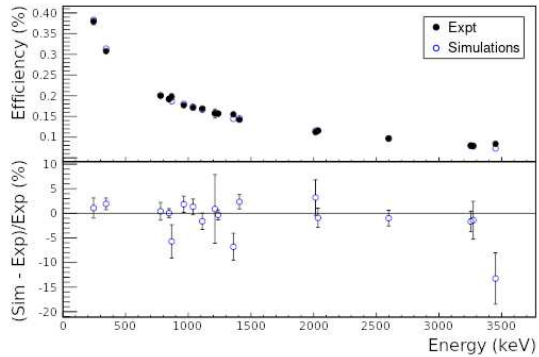}
    }
    \caption{\label{fig:Q4C4_outer} (Top) Measured (full circles) and
      simulated (open circles) photopeak efficiencies of a type-A
      crystal in a GRETINA quad using the optimal effective passive Ge
      layer thicknesses in the simulation including a 0.4~mm outer
      passive layer. (Bottom) Relative discrepancies between simulated
      and measured efficiencies.}
  \end{center}
\end{figure}

With back and coaxial effective passive-layer thicknesses determined
as described above, the average relative discrepancy between
simulations and measured photopeak efficiencies between 245~keV
and 3548~keV is 12\%. 
We investigated three explanations for this discrepancy.
First, we considered the possibility that passive material, external
to the crystal and between the crystal and the source, is missing
from the model. An additional thickness of about 6 mm of aluminum is
required to reduce the photopeak efficiency at 1~MeV by 12\%. We are
able to rule out this possibility, because the
aluminum beam tube, cryostat walls, and crystal encapsulation 
present a total thickness of 3.7~mm of aluminum, while the thickness of
the beam tube is known $\pm$0.25~mm, and the dimensions of the
capsules and walls are known within 0.1~mm tolerances.

Second, we applied a scaling to the 
polyhedra shaping the forward sides and front surfaces of the crystals
to reduce the overall crystal volume. We found that a scaling of 96\%
gave the best fit to measured photopeak efficiencies. This scaling
corresponds to a reduction in total crystal volume of 5.3\%. We rule
out this possibility, because the resulting crystal dimensions are not
compatible with the dimensions and tolerances in drawings provided by
the manufacturer, nor is the resulting total crystal volume consistent
with the crystal mass reported by the manufacturer.

Our third approach to accounting for the observed discrepancy between
simulated and observed photopeak efficiencies is the inclusion in the
crystal model of an outer effective passive layer of uniform thickness. 
We simulated photopeak efficiencies assuming outer passive-layer
thicknesses from 0.3~mm to 1.5~mm in 0.2~mm steps. The resulting
simulated photopeak efficiencies are shown as open circles in the
upper panel of Figure~\ref{fig:Q4C4_outer}. We found, by $\chi^2$
minimization, a best-fit outer effective passive-layer thickness of
0.40~mm with a 95\% confidence interval of 0.06~mm. This is a purely
statistical uncertainty within the limited parameter space of the
model and does not reflect the precision with which passive-layer
thicknesses can be determined.

\section{The Full Array}
\label{sec:full}

\subsection{Optimal Effective Passive Germanium Layers}
\label{sec:passive_full}

\begin{table}[h]
  \begin{center}
    \begin{tabular}{rccc}  \hline\hline
      & Minimum (mm) & Maximum (mm) & Step (mm) \\ \hline
      Back    & 1.0 & 6.0 & 0.25 \\ 
      Coaxial & 0.0 & 5.0 & 0.25 \\
      Outer   & 0.0 & 1.2 & 0.1  \\\hline\hline
    \end{tabular}
    \caption{\label{tab:grid} Ranges and step sizes of the
      three-dimensional grid of effective passive-layer thicknesses
      covered by the simulations of photopeak efficiencies of
      the full array.} 
  \end{center}
\end{table}

We used a similar heuristic approach to determine average effective
passive-layer thicknesses for the full 12-quad standard NSCL
configuration of GRETINA. We used the same strategy to constrain the
effective back passive-layer thickness for the full array as we did for
the single crystal.  However, pencil-beam scans have not been
collected for the entire array. Instead, we used measured photopeak
efficiencies for the full array to constrain both the effective coaxial
and outer passive Ge layer thicknesses. We explored the three-dimensional
parameter space in a grid specified in Table~\ref{tab:grid},
simulating photopeak efficiencies at 36 energies covering the range
53~keV to 3548~keV and calculating $\chi^2$ per degree of freedom for
both the full array and for events involving the back slices of the
crystals at each of the 5733 grid points. The $\chi^2/\mathrm{d.o.f.}$
values for events involving the back slices of the crystals were
calculated using the same 17 energies in the range 344~keV to
3548~keV used to constrain the back passive layer of a type-A crystal
in a GRETINA quad in Section~\ref{sec:back_outer_single}. 

We pared down the grid points by first selecting the back
passive-layer thickness giving the lowest $\chi^2$ for events
involving the back slices by fitting a cubic function to the $\chi^2$
vs. back passive-layer thickness results and identifying the best
thickness within the 0.25~mm precision of the grid.
We then further refined the search by
similarly selecting the outer passive-layer thickness within the
0.1~mm grid precision giving the lowest
$\chi^2$ for photopeak efficiencies for each of the remaining coaxial
and back passive-layer pairs.
We completed this process constraining the outer passive-layer
thickness using photopeak efficiencies in two energy ranges -- 
(I) 58~keV - 3548~keV and (II) 245~keV - 3548~keV. The reason for
selecting these energy ranges will become clear below, as we present
results of simulations performed with different sets of effective
passive-layer thicknesses which show varying degrees of success in
reproducing observed aspects of the measured spectra below 200~keV.

\begin{figure}
  \begin{center}
    \scalebox{0.55}{
      \includegraphics{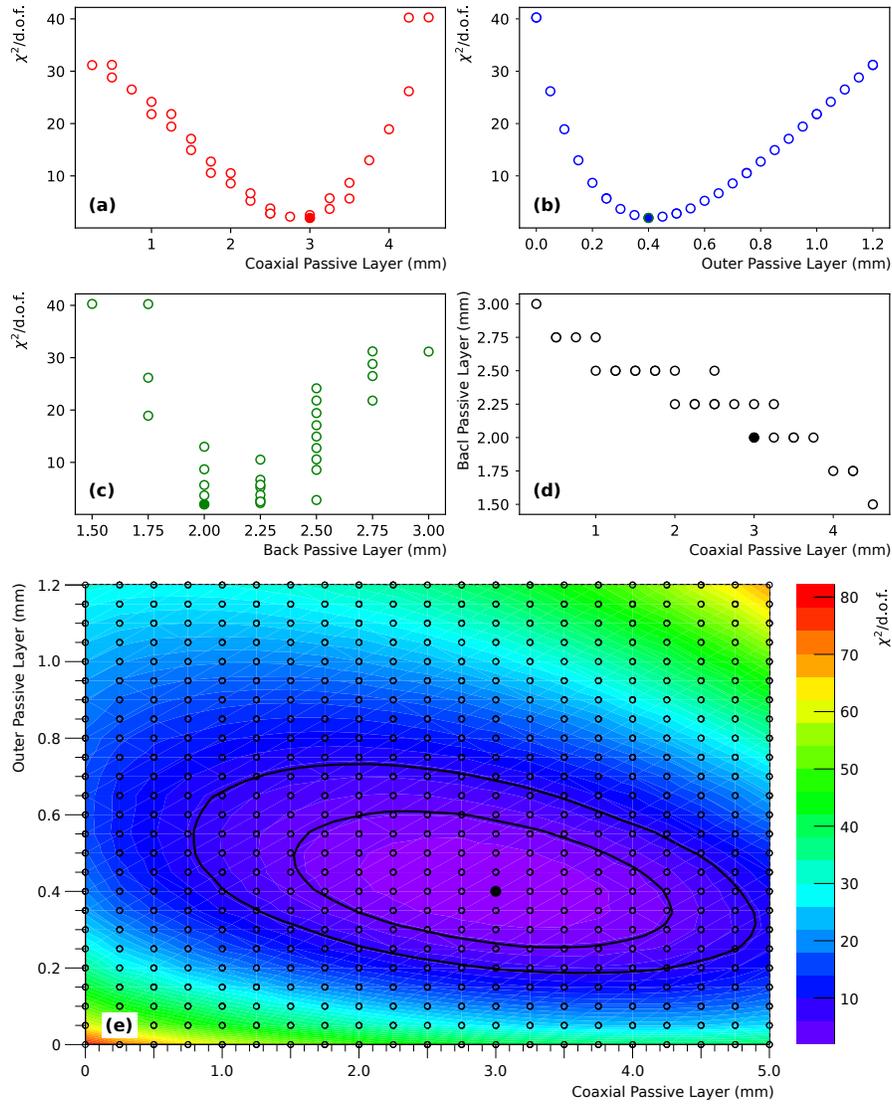}
    }
    \caption{\label{fig:layers} Plots of $\chi^2/\mathrm{d.o.f.}$
      vs. (a) coaxial, (b) outer, and (c) back effective passive-layer
      thicknesses, and (d) coaxial vs. outer passive-layer thickness
      for the grid points in the $\chi^2$ valley
      in the three-dimensional parameter space. The $\chi^2$ values
      were computed using the full set of efficiencies in energy range
      I between 53~keV and 3548~keV. The filled circles
      correspond to the optimal passive-layer thicknesses identified
      using the heuristic method described in the text. (e) The
      $\chi^2$ surface in the outer vs. coaxial passive-layer
      thickness plane at 2.0 mm back passive-layer thickness. The
      solid curves are the 68\% and 95\% confidence contours. The
      color scale and contours are determined by interpolations
      between grid points.}
  \end{center}
\end{figure}

\begin{figure}
  \begin{center}
    \scalebox{0.55}{
      \includegraphics{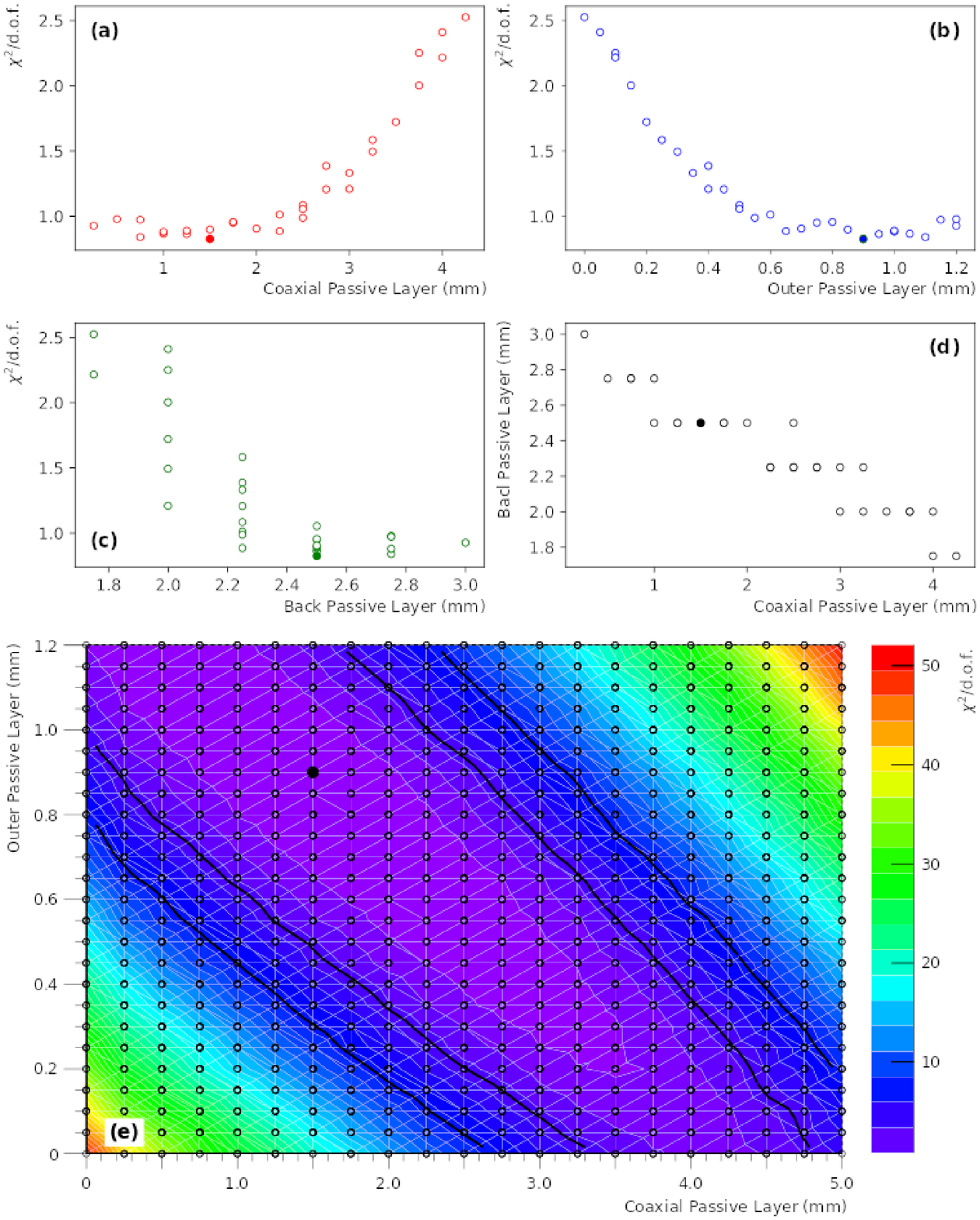}
    }
    \caption{\label{fig:layers_highE} Plots of
      $\chi^2/\mathrm{d.o.f.}$ vs. (a) coaxial, (b) outer, and (c)
      back effective passive-layer thicknesses, and (d) coaxial vs. outer
      passive-layer thickness for the grid points in the $\chi^2$
      valley in the three-dimensional parameter space. The $\chi^2$
      values were computed using efficiencies in energy range II
      between 245~keV and 3548~keV. The filled circles correspond to
      the optimal passive-layer thicknesses identified using the
      heuristic method described in the text. (e) The $\chi^2$ surface
      in the outer vs. coaxial passive-layer thickness plane at 2.5 mm
      back passive-layer thickness. The solid curves are the 68\% and
      95\% confidence contours. The color scale and contours are
      determined by interpolations between grid points.}
  \end{center}
\end{figure}

In Figures~\ref{fig:layers} and~\ref{fig:layers_highE}, panels (a)-(c) show
the surviving grid points, characterized by minimum $\chi^2$ per
degree of freedom from the cubic fit.  Panel (d) reveals the roughly
linear relationship between the back and coaxial passive-layer
thicknesses along a ``valley'' in the $\chi^2/\mathrm{d.o.f.}$ volume.
Panel (e) of Figures~\ref{fig:layers} and~\ref{fig:layers_highE} shows
$\chi^2/\mathrm{d.o.f.}$ surface in the plane corresponding to the
optimal effective back passive-layer thickness. The 68\% and 95\% confidence
contours are shown as solid curves, and the minimum of the surface is
marked by a filled circle. The color scales and contours are determined by
interpolations between grid points. The 95\% confidence interval for the back
passive-layer thicknesses in both cases is $\pm$1~mm. The two
resulting sets of optimal effective passive-layer thicknesses for the
full array are shown as filled circles in Figures~\ref{fig:layers}
and~\ref{fig:layers_highE} and are listed, along with those determined
for a type-A crystal in a GRETINA quad, in Table~\ref{tab:passivelayers}.

A comparison of Figures~\ref{fig:layers} and~\ref{fig:layers_highE}
reveals that the inclusion of the measured photopeak efficiencies at
energies below 245~keV constrains the outer passive layer much more
strongly. 

\begin{table}
  \begin{center}
    \begin{tabular}{ccccc}  \hline\hline
                & Energy Range & Coaxial (mm) & Back (mm) & Outer (mm) \\ \hline
      Type A     & I & 2.09 & 2.8 & 0.40 \\ \hline
      Full Array & I & 3.0  & 2.0 & 0.40 \\
                 & II & 1.5  & 2.5 & 0.90 \\\hline\hline
    \end{tabular}
    \caption{\label{tab:passivelayers} Sets of optimal effective
      passive-layer thicknesses for a type-A crystal in a GRETINA quad
      4, crystal 4 (Type A) and the full array determined using
      measured photopeak efficiencies in the energy ranges 53~keV -
      3548~keV (I) and 245~keV - 3548~keV (II).}
  \end{center}
\end{table}

\subsection{Photopeak Efficiencies}
\label{sec:photopeak_eff}

\begin{figure}
  \begin{center}
    \begin{tabular}{cc}
      \scalebox{0.40}{
        \includegraphics{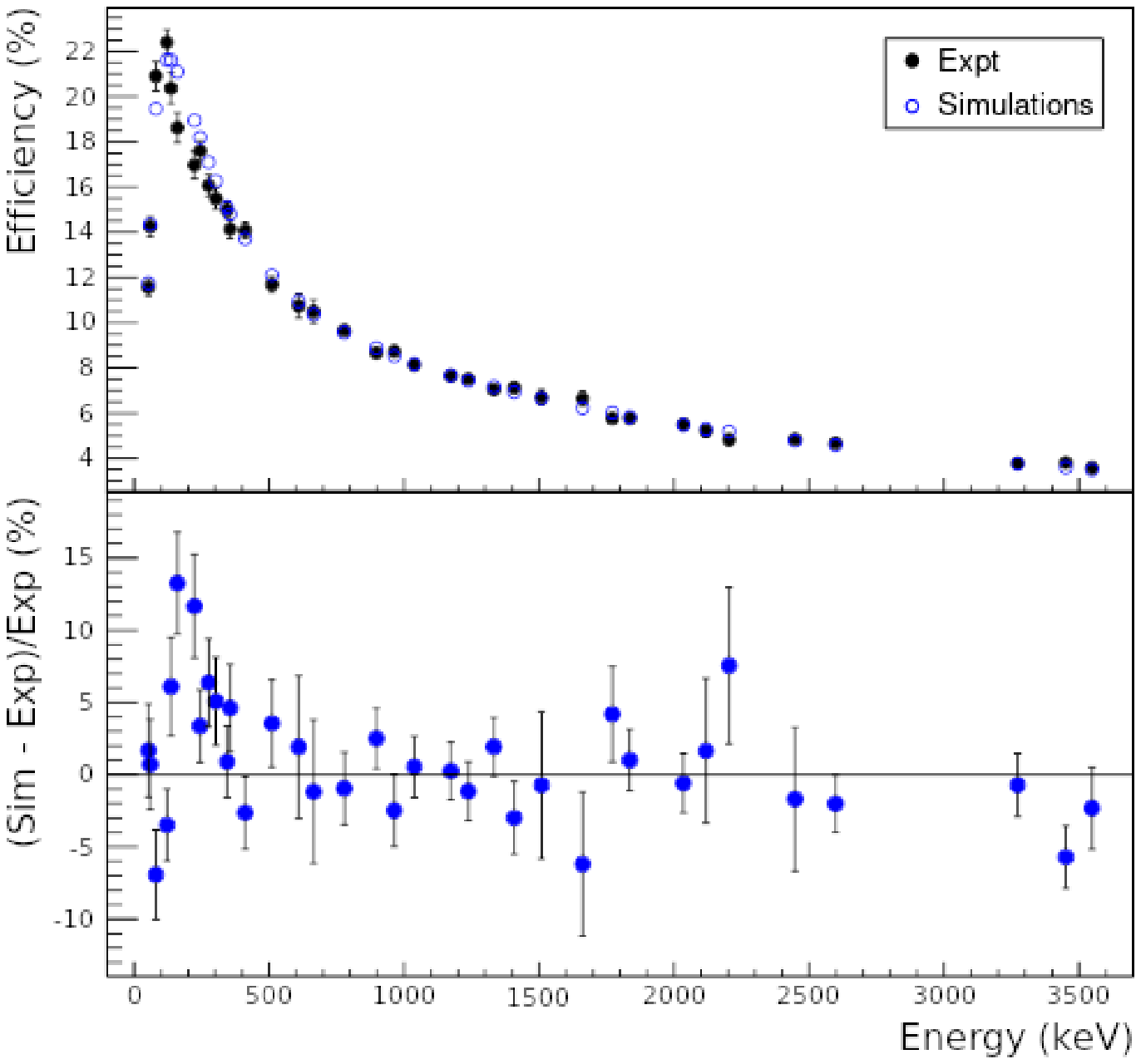}
      }
      &
      \scalebox{0.40}{
        \includegraphics{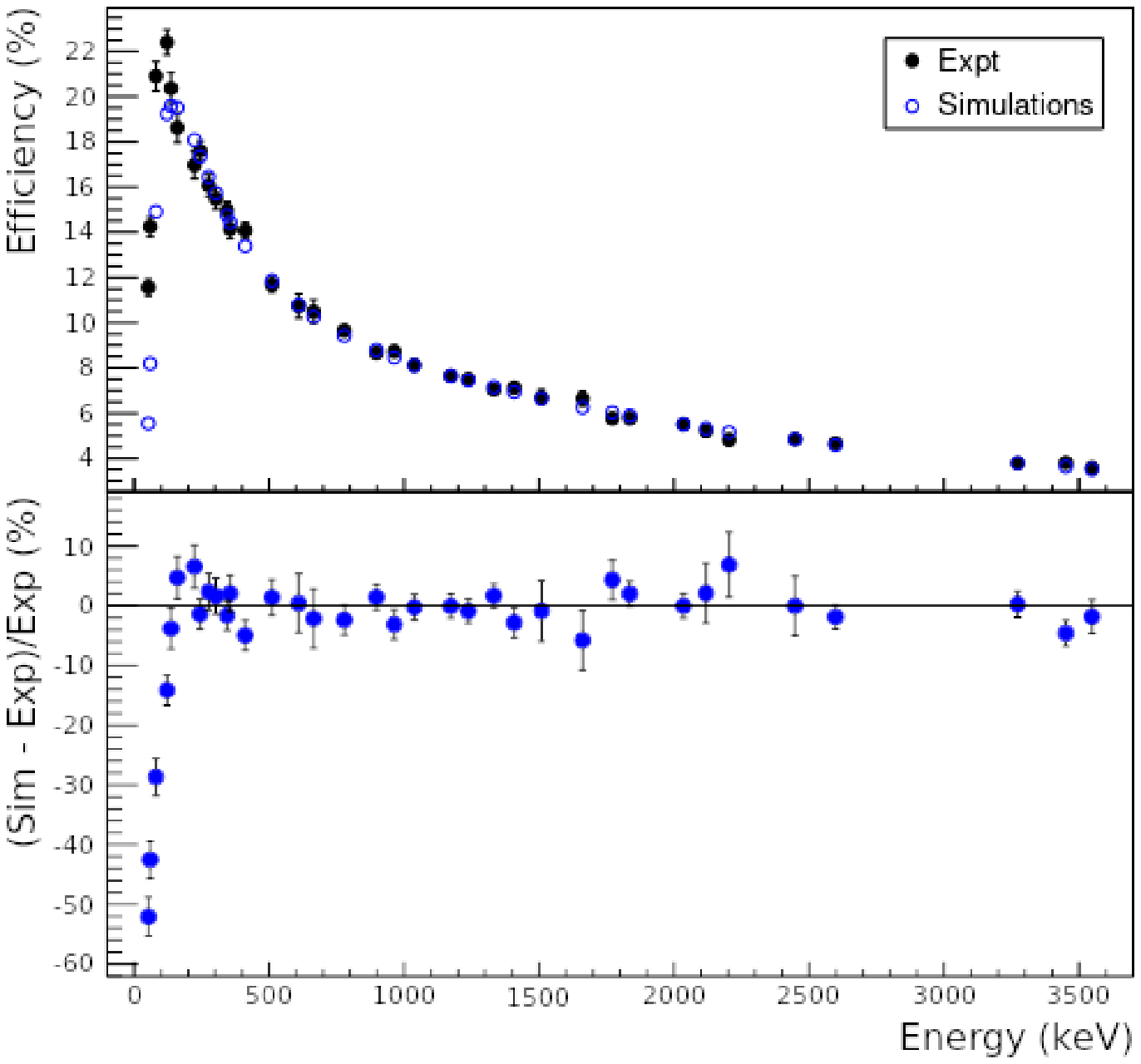}
      }\\
      {\fontfamily{\sfdefault}\selectfont \textbf{(a)}} &
      {\fontfamily{\sfdefault}\selectfont \textbf{(b)}}
    \end{tabular}
    \caption{\label{fig:eff_full_new} (Top panels) Measured (full
      circles) and simulated (open circles) photopeak
      efficiencies of the 12-quad standard NSCL configuration of the
      GRETINA array with the improved simulation model and 
      (bottom panels) Relative discrepancies between simulated and
      simulated and measured efficiencies with (a) 3.0~mm coaxial,
      2.0~mm back, and 0.40~mm effective passive layers determined
      using photopeak efficiencies in energy range I (53~keV -
      3548~keV) and (b) 1.5~mm coaxial, 2.5~mm back, and 0.90~mm outer
      effective passive layers determined using photopeak efficiencies
      in energy range II (245~keV - 3548~keV).}
  \end{center}
\end{figure}

\begin{figure}
  \begin{center}
    \scalebox{0.5}{
      \includegraphics{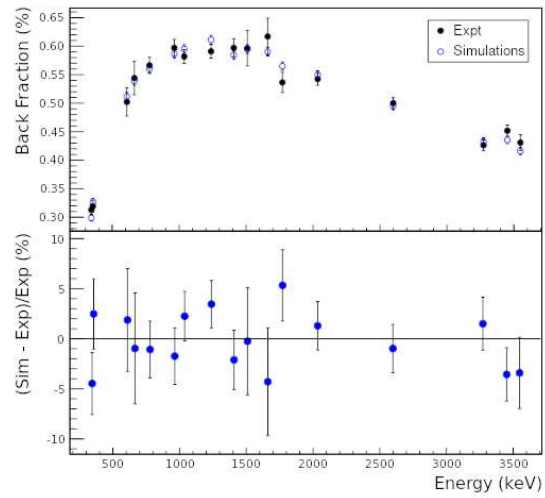}
    }
    \caption{\label{fig:full_back} (Top) Measured (full circles) and
      simulated (open circles) fraction of photopeak events in the
      12-quad standard NSCL configuration of GRETINA with at
      least one interaction point in the back slice of the crystals
      using the optimal effective passive-layer thicknesses determined
      using the wider energy range I in the simulations. (Bottom) Relative
      discrepancies between simulated and simulations and
      measurements.}
  \end{center}
\end{figure}

In Figure~\ref{fig:eff_full_new}, 
measured photopeak efficiencies of the 12-quad standard NSCL
configuration of the GRETINA array covering a broad range of
gamma-ray energies are compared with simulations using the two sets of
effective passive-layer thicknesses described in
Section~\ref{sec:passive_full}. The simulations used the full model of
GRETINA and surrounding passive material shown in
Figure~\ref{fig:ucg_full}(b). Measured and simulated back fractions
for the full array, using passive-layer set I, are shown in
Figure~\ref{fig:full_back}. 

\subsection{Compton Continuum}
\label{sec:Compton}

\begin{figure}
  \begin{center}
    \scalebox{0.5}{
      \includegraphics{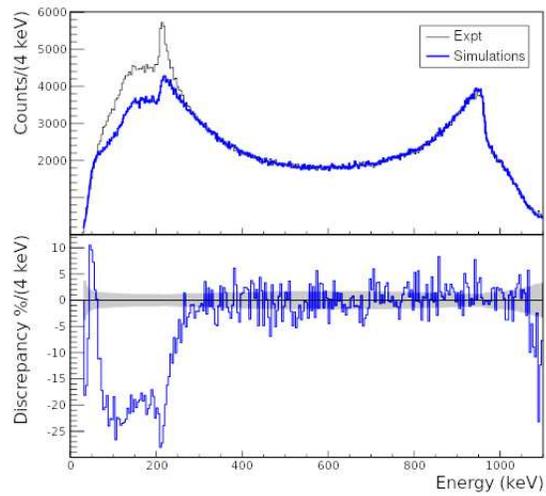}
    }
    \caption{\label{fig:Compton_noextra} (Top) Measured (thin black
      line) and simulated (heavy blue line) Compton continuum of the 
      1172~keV gamma ray in $^{60}$Ni. The simulation used the minimal
      model of GRETINA in Figure~\ref{fig:ucg_full}(a) and the optimal
      effective passive-layer thicknesses determined using photopeak
      efficiencies in energy range I.
      (Bottom) Relative residual spectrum. The shaded region
      corresponds to the statistical uncertainty in each bin. It is
      evident that back scattering from material around the Ge shell,
      like the aluminum mounting shell and detector cryostats, need to
      be included in the model. }

  \end{center}
\end{figure}

\begin{figure}
  \begin{center}
    \begin{tabular}{cc}
      \scalebox{0.38}{
        \includegraphics{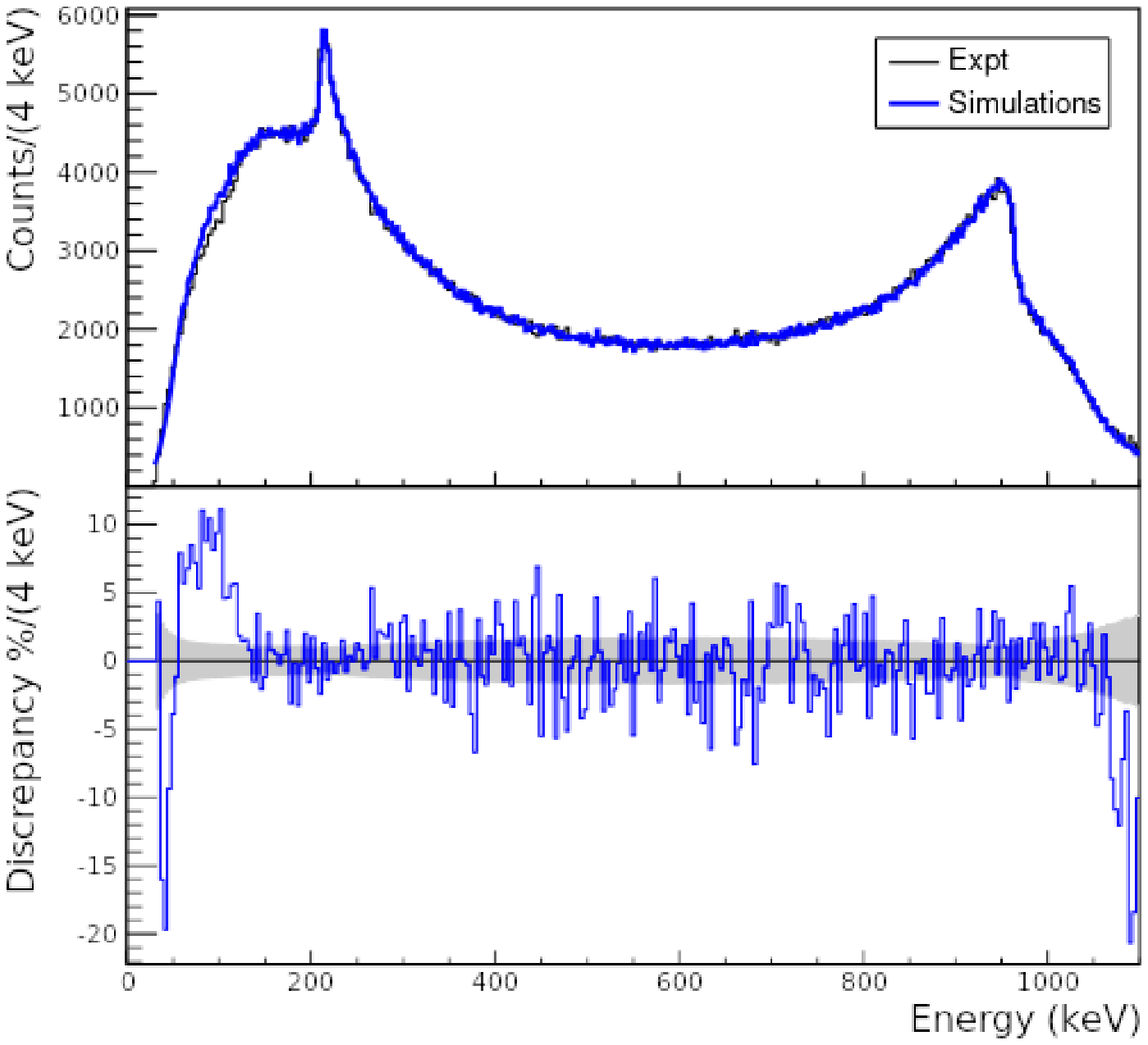}
      }
      &
      \scalebox{0.38}{
        \includegraphics{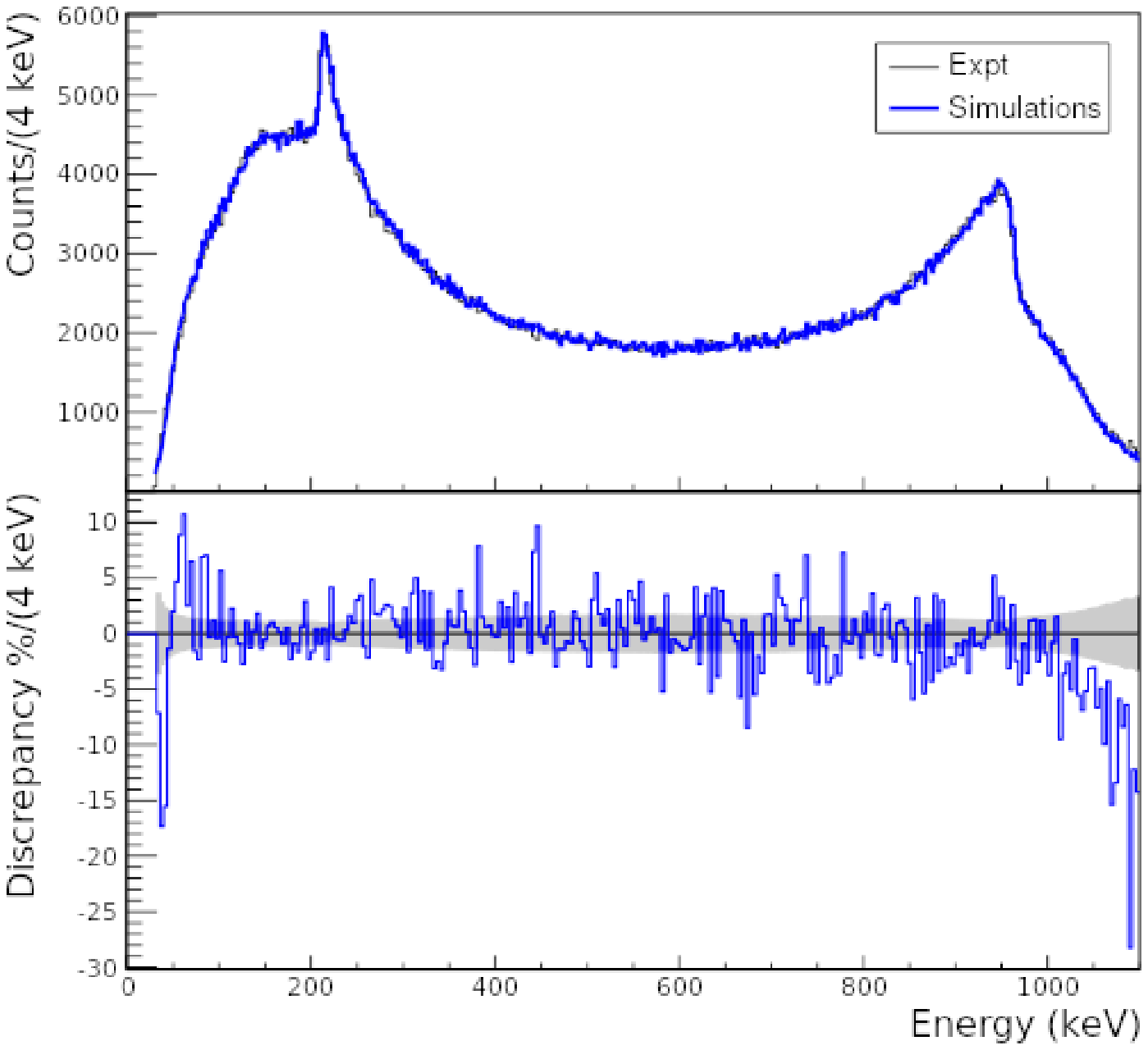}
      }\\
      {\fontfamily{\sfdefault}\selectfont \textbf{(a)}} &
      {\fontfamily{\sfdefault}\selectfont \textbf{(b)}}
    \end{tabular}
    \caption{\label{fig:Compton} (Top panels) Measured (thin black
      line) and simulated (heavy blue line) Compton continuum of the
      1172~keV gamma ray in $^{60}$Ni. (Bottom panels) Relative
      residual spectrum. The regions shaded in gray correspond to the 
      statistical uncertainty in each bin. The simulations used the
      full model of GRETINA shown in Figure~\ref{fig:ucg_full} and 
      effective passive-layer thicknesses determined using (a)
      photopeak efficiencies in energy range I (53~keV - 3548~keV) and
      (b) photopeak efficiencies in energy range II (245~keV -
      3548~keV).} 
  \end{center}
\end{figure}

In Figure~\ref{fig:Compton_noextra}, the measured Compton continuum of
the 1172~keV gamma ray in $^{60}$Ni is compared with simulations using
the model including minimal additional passive material illustrated in
Figure~\ref{fig:ucg_full}(a). The simulations in
Figure~\ref{fig:Compton} used the full model illustrated in
Figure~\ref{fig:ucg_full}(b).  Figure~\ref{fig:Compton}(a) used
effective passive-layer thicknesses determined using photopeak
efficiencies in energy range I, and Figure~\ref{fig:Compton}(b) used
effective passive-layer thicknesses determined using photopeak
efficiencies in energy range II.  In both cases, the spectrum below
30~keV is excluded from the fit due to variations in low-energy
thresholds among the crystals, not included in the simulated spectra.
The shaded regions in the residual spectra in the bottom panels of
Figure~\ref{fig:Compton_noextra} correspond to the statistical
uncertainties in each bin of the measured and simulated spectra added
in quadrature.

Figure~\ref{fig:Compton_noextra} reveals large discrepancies
between the measured and simulated spectra in the portion
of the Compton continuum below 300~keV. The inclusion of significant
passive material surrounding the active detector volumes, but not in
direct line of sight with the source, is needed in order to  
obtain the improved agreement shown in Figure~\ref{fig:Compton}. The
discrepancies below 150~keV in Figure~\ref{fig:Compton}
correspond roughly to the discrepancies in low-energy photopeak
efficiencies evident in Figure~\ref{fig:eff_full_new}. 
We tentatively attribute the statistically-significant discrepancies
in the region above 1000~keV, approaching the photopeak, to
partial charge collection of full-energy events.

The low energy region of the Compton continuum below 150 keV in
Figure~\ref{fig:Compton} clearly shows that the simulation constrained
by the wider energy range I, and resulting in a thinner outer effective
passive layer over-predicts the spectrum counts by up to 10\%. The
simulation using the thicker dead layer provides a much better
description in that region. Both simulations describe well the Compton
continuum above 150~keV.

\section{Summary}

We have described the UCGretina \textsc{geant4} simulation code and
the development of an accurate model of the array, for use in the
planning and analysis of Doppler-reconstructed gamma-ray spectra
collected with GRETINA in in-beam gamma-ray spectroscopy measurements
with beams traveling at $v/c \gtrsim 0.3$ at the NSCL and FRIB. 
We determined that the inclusion of millimeter-scale passive layers
surrounding the active volumes of the crystals, including an outer
passive layer, and additional passive material behind and adjacent to
the active detector volumes, yielded significant improvements in the
agreement between simulated and measured photopeak efficiencies and
the Compton continuum.  We have
presented heuristic methods for determining optimal effective
passive-layer thicknesses for a single crystal and average thicknesses
for the full array.

Despite the success of the simple model of passive Ge layers adopted
here in reproducing the overall detector response, it should not be
understood as a realistic physical representation of the regions near
the crystal surfaces. In that regard, we use the term
\textit{effective} in reference to the passive layers in the model
throughout the present work.

An important finding is that particular care must be taken with the
choice of the thicknesses of the effective passive-layers in
simulations of gamma rays at energies below 250~keV depending on the
aspect of the spectrum which is intended to be modeled. The 0.40~mm
outer passive-layer thickness determined using the full range of
measured photopeak efficiencies, from 58~keV to 3549~keV, leads to a
simulation describing the gamma-ray efficiency well for energies above
300~keV and below 100~keV, but over-predicts systematically
efficiencies by up to 15\% in the energy range from
100-300~keV. Furthermore, this model struggles to describe the Compton
continuum stemming from a 1173~keV gamma ray in the same energy
range. Determining the passive-layer thicknesses considering only
energies $\geq$245~keV more than doubles the thickness of the outer
effective passive layer for optimal description in the considered
energy range, 
which surprisingly extends the energy range of good agreement with
measured photopeak efficiencies and the Compton continuum down to
100~keV, though the thicker outer passive layer significantly
under-predicts the measured efficiencies below 100~keV. These results
are compatible with Ref.~\cite{Mai16}, 
showing that the effective outer passive-layer geometry of a coaxial
HPGe detector is energy dependent. Our results suggest that
energy-dependent effects become significant in GRETINA crystals at
gamma-ray energies below approximately 100~keV.

Prospects for improving on the average crystal model presented here
include accounting for variations in dimensions and passive-layer 
thicknesses among crystals. The discrepancies between optimal
coaxial and back passive-layer thicknesses determined for a single
crystal and for the full array indicate that these variations may be
significant. In addition, it is important for applications such as
polarimetry and gamma-ray tracking to model realistic segmentation and
partial charge collection for interaction points near boundaries
between passive layers and active detector volumes.

\section{Acknowledgments}

This work was supported by the NSF under Grant Nos. PHY-1303480,
PHY-1617250, and PHY-1102511 and by the Department of Energy, Office
of Science, under Grant No. DE-SC0020451 (MSU).  GRETINA was funded by
the DOE, Office of Science. Operation of the array at the NSCL was
supported by the DOE under Grants No. DE-SC0014537 and DE-SC0019034
(NSCL) and No. DE-AC02-05CH11231 (LBNL).
We are grateful to A. O. Macchiavelli for productive conversations
throughout the project. We also thank T. J. Carroll for the use of the
Ursinus College Parallel Computing Cluster, supported by the NSF under
Grant No. PHY-1607335.

%\section*{References}

%\bibliography{ucgretina}

\end{document}